\documentclass[epj,final]{svjour}
\usepackage{graphics}
\input{epsf} 

\begin{document}
\title{Compressibility crossover and quantum opening of a gap 
for two dimensional disordered clusters with Coulomb repulsion} 
\author{Giuliano Benenti\inst{1}, Xavier Waintal\inst{1}, 
Jean-Louis Pichard\inst{1} \and Dima L. Shepelyansky\inst{2}}  
\institute{CEA, Service de Physique de l'Etat Condens\'e, 
Centre d'Etudes de Saclay, 91191 Gif-sur-Yvette, France  
\and 
Laboratoire de Physique Quantique, UMR C5626
du CNRS, Universit\'e Paul Sabatier, 31062 Toulouse, France} 
\titlerunning{Compressibility crossover and quantum opening 
of a gap in $2d$} 
\authorrunning{G. Benenti {\it et al.}} 
\date{Received:}
\abstract{ 
 Using Hartree-Fock orbitals with residual Coulomb repulsion, 
we study spinless fermions in a two dimensional random potential. 
When we increase the system size $L$ at fixed particle density, the 
size dependence of the average inverse compressibility exhibits a 
smooth crossover from a $1/L^2$ towards a $1/L$ decay when the 
Coulomb energy to Fermi energy ratio $r_s$ increases from $0$ to $3$. 
In contrast, the distribution of the first energy excitation displays 
a sharp Poisson-Wigner-like transition at $r_s \approx 1$. 
}

\maketitle

\section{Introduction} 
\label{intro} 

 The interplay of disorder and interactions in two dimensional ($2d$) 
electronic systems is currently a central problem in condensed 
matter physics \cite{loc99}. The importance of interactions is 
illustrated by several phenomena, notably the discovery of a metallic 
phase in various two-dimensional devices \cite{Kravchenko}. Since 
the metal-insulator transition occurs at a Coulomb energy to Fermi 
energy ratio $r_s \approx 10$, Coulomb repulsion should be crucial 
in understanding this phenomenon. Compressibility measurements 
indicate that the high $r_s$ insulating phase is incompressible 
\cite{comp1} and spatially inhomogeneous \cite{comp2}. 

 In weakly disordered quantum dots, the inverse compressibility gives 
the spacing between adjacent conductance peaks. The peak spacing 
statistics, obtained in experiments in which $r_s\approx 1$, 
displays fluctuations larger than those predicted from Random 
Matrix Theory, characterized by a Gaussian-like distribution instead 
of a Wigner-Dyson distribution \cite{Sivan,Marcus,Simmel}. 
This suggests a breakdown of the naive single particle picture, 
assumed in the constant interaction model. 
  
 In strongly disordered insulators, the long range nature of interactions  
leads to a soft Coulomb gap in the single particle density of 
states at the Fermi level \cite{ES}, observed in tunneling 
spectroscopic experiments in $3d$ \cite{Lee} and $2d$ 
\cite{Ashoori} nonmetallic semiconductors.    
Assuming that single electron hoppings dominate the transport, 
the Coulomb gap leads to a crossover in the temperature dependence 
of the resistivity $\rho(T)$ from the Mott variable range hopping 
law ($\rho(T)=\rho_M\exp(T_0/T)^{1/3}$ in dimension $d=2$) to the 
Efros-Shklovskii behavior ($\rho(T)=\rho_{ES}\exp(T_{ES}/T)^{1/2}$)
\cite{ES}. This crossover has been reported \cite{khondaker} not 
only decreasing temperature but also decreasing carrier density 
(around $r_s\approx 1.7$).    

 In this paper, we numerically investigate the physics of $2d$ 
Coulomb interacting spinless fermions in a random potential for 
relatively low values of the factor $r_s$ ($\leq 3$), when 
the single particle spectra display either Wigner-Dyson statistics 
for weak disorder (diffusion) or Poisson statistics for strong  
disorder (Anderson localization). As 
exact diagonalization studies are restricted to small system sizes 
\cite{paper1}, we use a numerical method \cite{CIphys1,CIphys2} 
familiar in quantum chemistry as the configuration interaction 
method (CIM) \cite{CIchem}. The method consists in diagonalizing 
the Hamiltonian in an energetically truncated basis built of  
the low-energy states of the corresponding Hartree-Fock (HF) 
Hamiltonian. This method allows us to study the ground state and 
the lowest energy excitations for different system sizes $L$ at  
constant electronic density.    

 We show that the average inverse compressibility $\Delta_2$ exhibits 
a smooth crossover from a $1/L^2$ towards a $1/L$ decay when $r_s$ 
increases from $0$ to $3$. At the same time, the distribution of 
$\Delta_2$ evolves from the Wigner-Dyson distribution (or the Poisson 
distribution, if disorder is strong enough or $L$ large enough) 
towards a Gaussian-like shape. 
Therefore fluctuations in $\Delta_2$ are determined by Coulomb 
repulsion rather than from single particle level fluctuations.  

 We have also studied the energy level spacing between the ground 
state and the first excitation for strongly disordered clusters: 
the average indicates a smooth opening 
of a gap when $r_s$ increases, while the distribution exhibits 
a sharp Poisson-Wigner-like transition at $r_s=r_s^C\approx 1.2$.  
The critical threshold $r_s^C$ is characterized by a scale 
invariant gap distribution, reminiscent of the one particle problem 
\cite{Shklovskii} at a mobility edge. However, it is only the 
distribution of the first spacing which exhibits such a transition, the 
distributions of the next spacings remain Poissonian and
are essentially unchanged when $r_s$ varies.

 The paper is organized as follows: in Section \ref{model} we 
introduce the model, in Section \ref{cim} we discuss the numerical 
technique used and its limits of validity;  
in Section \ref{comp} we present our compressibility data; in 
Section \ref{gap} we discuss the statistical properties of 
the lowest energy excitations; in Section \ref{hop} we discuss 
the implications of our results for hopping conductivity 
experiments; in Section \ref{conc} we present our conclusions.    
 
\section{The model} 
\label{model} 

 We consider a disordered square lattice with $M=L^2$ sites occupied by 
$N$ spinless fermions. The Hamiltonian reads 
\begin{equation} 
\label{hamiltonian} 
H=-t\sum_{<i,j>} c^{\dagger}_i c_j +  
\sum_i v_i n_i  + U \sum_{i\neq j} \frac{n_i n_j } {2r_{ij}},
\end{equation} 
where $c^{\dagger}_i$ ($c_i$) creates (destroys) an electron in 
the site $i$, the hopping term $t$ between nearest neighbours 
characterizes the kinetic energy, $v_i$ the site potentials 
taken at random inside the interval $[-W/2,+W/2]$, 
$n_i=c^{\dagger}_i c_i$ is the occupation number at site $i$ and $U$ 
measures the strength of the Coulomb repulsion. The boundary conditions 
are periodic and $r_{ij}$ is the inter-particle distance for a $2d$ torus. 
If $a^*_B=\hbar^2\epsilon/(m^* e^2)$, $m^*$, $\epsilon$, $a$ and 
$n_s=N/(a L^2)$ denote respectively the effective Bohr radius, 
the effective mass, the dielectric constant, the lattice spacing  
and the carrier density, the factor $r_s$ is given by:
\begin{equation}
r_s=\frac{1}{\sqrt{\pi n_s} a^*_B} = \frac{U}{2t\sqrt{\pi n_e}}, 
\end{equation} 
since in our units $\hbar^2/(2m^*a^2)\to t$, $e^2/(\epsilon a)\to U$ 
and $n_e=N/L^2$. 

\section{Configuration interaction method}
\label{cim} 

 A numerical study of the model (\ref{hamiltonian}) via exact 
diagonalization techniques for sparse matrices is possible only for 
small systems \cite{paper1}, and does not allow us to vary $L$ for a 
constant density. We are obliged to look for an 
approximate solution of the problem, using the Hartree-Fock orbitals, 
and to control the validity of the approximations. One starts from the 
HF Hamiltonian where the two-body part is reduced to an effective single 
particle Hamiltonian 
\cite{Kato,Poilblanc,Schreiber}
\begin{equation} 
\label{hfhamiltonian} 
\begin{array}{c}
\displaystyle{ 
U (\sum_{i\neq j} \frac{1}{r_{ij}} n_i\langle n_j \rangle
- \sum_{i\neq j} \frac{1}{r_{ij}} c^{\dagger}_i c_j  
\langle c^{\dagger}_j c_i \rangle }),  
\label{HF}
\end{array} 
\end{equation} 
where $\langle ... \rangle$ stands for the expectation value with 
respect to the HF ground state, which has to be determined 
self-consistently. For large values of the interaction 
and large system sizes the single-particle problem (\ref{hfhamiltonian}) 
is still non-trivial, since the self-consistent iteration can be trapped 
in metastable states. This limits our study to small $r_s$ 
and forbids us to study by this method charge crystallization 
discussed in \cite{paper1} at a larger $r_s^W \approx 10$ .      

\begin{figure} 
\centerline{
\epsfxsize=8cm 
\epsfysize=8cm 
\epsffile{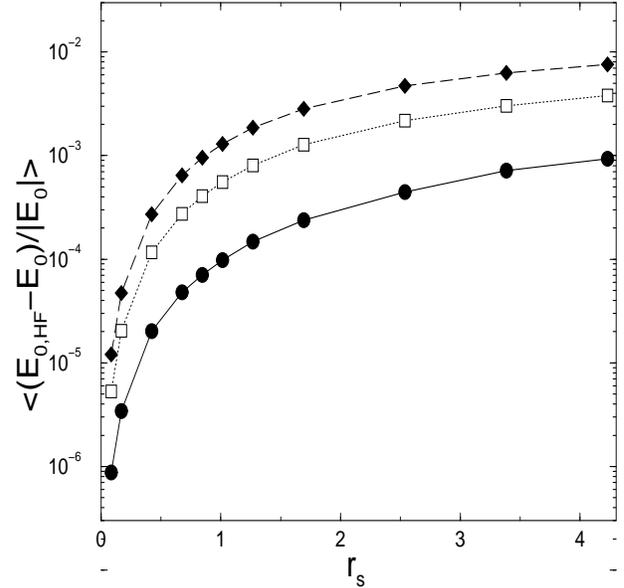}
}
\caption{Relative error in the HF ground state energy as a 
function of $r_s$, for $N=4$, $L=6$, $W=5$ (circles), $W=8$ 
(squares), and $W=15$ (diamonds). The contribution of a positive 
uniform background has been included in the energies. Brackets 
indicate disorder average (here over $10^2$ configurations).    
}
\label{fig1} 
\end{figure}

 Fig. \ref{fig1} shows that the HF approximation gives 
a good estimate of the ground state energy. This approximation 
becomes worse for (1) larger interactions or (2) smaller disorder. 
The first characteristic is due to the fact that the ground state HF 
energy is exact in the limits $U\to 0$ and $U \to \infty$, 
in which the ground state is given by a single Slater determinant, 
but deviations from this simple picture are expected 
at intermediate $U$ values. 
The latter feature can be understood as complicated many-body 
screening effects, which are effective up to a distance of the 
order of the localization length, cannot be reproduced within 
simple HF. 

 The main advantage of the HF approximation is that it reproduces 
well the single particle density of states\cite{Schreiber}, 
particularly the soft Coulomb gap at the Fermi energy \cite{ES}. 
However, the approximations involved in the HF method are 
uncontrolled. The mean field HF results can be improved using the 
configuration interaction method \cite{CIphys1,CIphys2,CIchem}. Once 
a complete orthonormal basis of HF orbitals has been calculated:  
\begin{equation} 
H_{HF}(|\psi_1\rangle,...,|\psi_N\rangle)  
|\psi_{\alpha}\rangle = \epsilon_{\alpha}|\psi_{\alpha}\rangle,    
\end{equation} 
with $ \alpha=1,2,\ldots,L^2$, it is possible to build up a Slater 
determinants' basis for the many-body problem which can be truncated to 
the $N_H$ first Slater determinants, ordered by increasing energies. 
The two-body Hamiltonian can be written as 
\begin{equation}
H_{\rm int}=\frac{1}{2} \sum_{\alpha,\beta,\gamma,\delta}  
Q_{\alpha\beta}^{\gamma\delta} d_{\alpha}^{\dagger} d_{\beta}^{\dagger} 
d_{\delta}d_{\gamma} ,
\label{Hint}
\end{equation} 
with 
\begin{equation} 
Q_{\alpha\beta}^{\gamma\delta} = 
U\sum_{i\neq j} \frac{ \psi_\alpha(i) \psi_\beta(j) \psi_\gamma(i) 
\psi_\delta (j)}{r_{ij}}  
\end{equation}
and $d^{\dagger}_{\alpha}=\sum_j \psi_{\alpha} (j) c^{\dagger}_j |0 \rangle$. 
One gets the residual interaction subtracting Eq. \ref{HF} from  
Eq. \ref{Hint}. This keeps the two-body nature of 
the Coulomb interaction, and if $N\gg 2$ it is still possible to take 
advantage of the sparsity of the matrix and to diagonalize it via the 
Lanczos algorithm.  

\begin{figure} 
\centerline{
\epsfxsize=8cm 
\epsfysize=8cm 
\epsffile{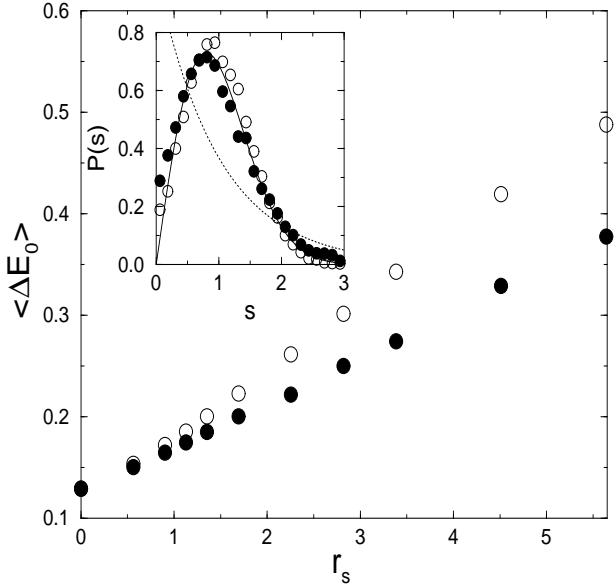}
}
\caption{ 
CIM result (filled circles) compared to HF approximation (empty circles)
for $N=9$, $L=12$, $W=15$. Mean gap $< \Delta E_0 >$ and gap distribution 
at $r_s=4.5$ (inset). Disorder average is over $10^4$ configurations. 
}
\label{fig2} 
\end{figure}

 Fig. \ref{fig2} compares HF and CIM results. Labelling the $N$-body 
levels $E_i$ ($i=0,1,2,...$) by increasing energy we have studied the 
first spacing $\Delta E_0=E_1-E_0$. 
When $r_s > 1$, the residual interaction significantly reduces the mean 
gap, and slightly changes its distribution. Within HF approximation,  
the first excitation is a particle-hole excitation starting from 
the ground state: The mean gap reduction is due to correlation effects 
beyond the particle-hole interaction. For example, the electron (hole)
polarizes its neighborhood as it creates fluctuations in the local 
charge density. 

\begin{figure}
\centerline{
\epsfxsize=8cm 
\epsfysize=8cm 
\epsffile{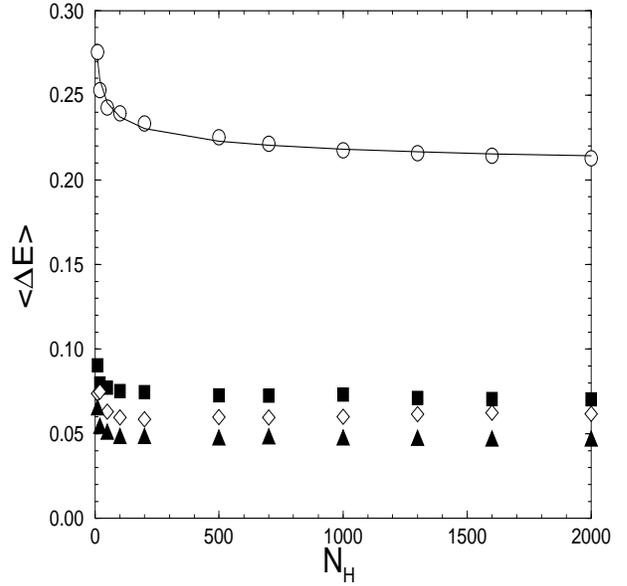}
}
\caption{First (circles), second (squares), third (diamonds), 
and fourth (triangles) average spacing as a function of 
the size $N_H$ of the truncated Hilbert space, for $N=16$, 
$L=16$, $W=15$, $r_s=2.8$. The first gap is fitted by the curve 
$<\Delta E_0>=0.195+0.13 N_H^{-0.26}$; therefore one can estimate 
the relative error at $N_H=2\times 10^3$ as $<10\%$. 
Disorder average is over $10^2$ configurations.  
}
\label{fig3} 
\end{figure}

 We have checked that CIM results agree with the results given from exact 
diagonalization with an accuracy of the order $2\%$ when one takes into 
account the $N_H=10^3$ lowest energy Slater determinants when $N=4$, $L=8$, 
$W=15$, and $r_s=5$. This means that a basis spanning only $0.2\%$ of 
the total Hilbert space is sufficient for studying the first excitations. 
For larger $L$, exact diagonalization is no longer possible, but one can 
look the variation of the results when $N_H$ increases. 
In the worst case considered ($N=16$, $L=16$, $W=15$, $r_s=2.8$) 
the accuracy in the first four spacings can be estimated as better   
than $10\%$ when $N_H=2\times 10^3$ (see Fig. \ref{fig3}). 

 Therefore the CIM method allows to study low energy level statistics
for $r_s<3$. However, its accuracy is not sufficient to determine more 
sensitive quantities, like a small change of the ground state energy 
when the boundary conditions are twisted (i.e. the persistent currents).
 
\section{Electronic compressibilty}
\label{comp} 

 In quantum dots experiments in the Coulomb blockade regime the 
spacing between consecutive conductance peaks is given by  
\begin{equation} 
\label{invcomp} 
\Delta_2(N)=E_0(N+1)-2E_0(N)+E_0(N-1),   
\end{equation} 
with $E_0(N)$ $N$-body ground state. This quantity is the discretized  
second derivative of the ground state energy with respect to the number 
of particles, i.e. the inverse compressibility.  
In the constant interaction model, which ignores fluctuations  
in the Coulomb energy 
\begin{equation}
Q_{\alpha\beta}^{\gamma\delta}=
\frac{e^2}{C} \delta_{\alpha\gamma}\delta_{\beta\delta},
\end{equation}
where $C$ is the capacitance of the dot. 
This gives $E_0(N)=e^2 N(N-1)/2 C +\sum_{k=1}^N \eta_k$, 
where $\eta_k$ is the $k$-th single particle energy at $U=0$. 
Substituting this expression into Eq. (\ref{invcomp}) one finds  
$\Delta_2=e^2/C+\eta_{N+1}-\eta_N$. 
Random Matrix Theory describes the single particle level spacings 
when the disorder is weak. The peak spacing distribution is 
expected to follow the Wigner-Dyson distribution with fluctuations 
in $\Delta_2$ whose standard deviation 
is given by $\delta \Delta_2\equiv \sqrt{<\Delta_2^2>-<\Delta_2>^2}= 
\sqrt{4/\pi-1}<\Delta>\approx 0.52<\Delta>$, where $<\Delta>$ is the single 
particle mean level spacing. However, experiments \cite{Sivan,Marcus,Simmel} 
performed at $r_s\approx 1$ found a distribution which 
is Gaussian-like and has a larger width, up to $7.5\Delta$ \cite{Simmel}.  
This suggests that fluctuations in $\Delta_2$ are dominated by 
electron-electron interactions. 
Several observed features of the peak spacing distributions have 
been reproduced using exact numerical diagonalization of the Hamiltonian 
model (\ref{hamiltonian}) \cite{Sivan}, HF calculations 
\cite{Levit,Walker,Berkovits}, and a random matrix model for 
interacting fermionic systems \cite{Alhassid}.  
 
 In this paper we study the inverse compressibility at different 
system sizes, for a constant filling factor $n_e=1/9$. 
We consider $N=4,9,16$ particles on square lattices of size 
$L=6,9,12$ respectively. A Fermi golden rule approximation for  
the elastic scattering time gives \cite{Walker}, for $n_e\ll 1$,  
$k_F l\approx 192 \pi n_e (t/W)^2$, $k_F$ and $l$ denoting the 
Fermi wave vector and the elastic mean free path, respectively.  
We consider $W=5,8,15$, corresponding to $k_F l\approx 2.7,1,0.3$; 
the first case corresponds to a diffusive system ($L>l/a\approx 2.3$, 
with $a$ lattice spacing), the last one to a strongly localized system.  

 The following inverse compressibility data are obtained with the CIM.   
We have checked that the residual interaction does not change 
qualitatively HF results. This is consistent with Fig. \ref{fig1}: 
HF approximation gives a good estimate of the ground state 
energy and the inverse compressibility is a physical observable which 
depends only on the ground state energies at different number of particles. 
Neither a precise knowledge of the ground state wavefuction (as in 
the calculation of persistent currents) nor excited states energies 
(as in studies of spectral statistics) are required. 

\begin{figure} 
\centerline{
\epsfxsize=8cm 
\epsfysize=8cm 
\epsffile{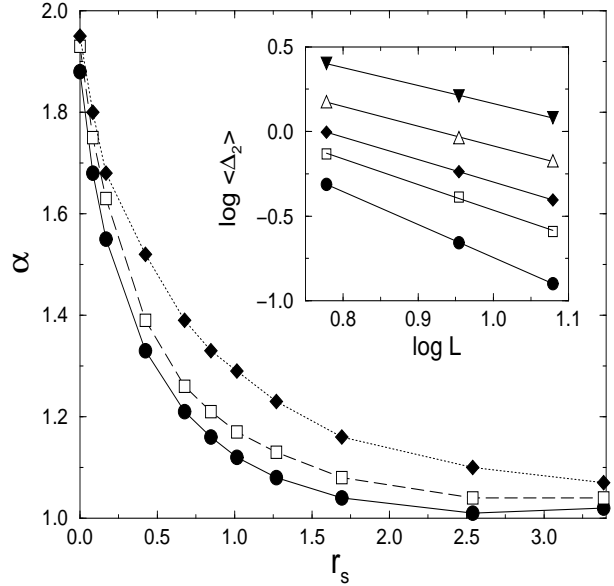}
}
\caption{ 
Inset: size dependence of the average inverse compressibility $<\Delta_2>$, 
for $W=15$, filling factor $n_e=1/9$, $r_s=0$ (circels), $0.4$ (squares), 
$0.8$ (diamonds), $1.7$ (triangles up), and $3.4$ (triangles down). 
Straight lines are power law fits $<\Delta_2(r_s)>\propto
L^{-\alpha(r_s)}$. 
Main figure: Exponent $\alpha(r_s)$ for $n_e=1/9$,  
$W=5$ (circles), $W=8$ (squares), and $W=15$ (diamonds). 
Disorder average for the compressibility data is over $10^3$ 
configurations.  
}
\label{fig4} 
\end{figure}

 Fig. \ref{fig4} shows the $L$-dependence of the average inverse 
compressibility, which is well fitted by the power law  
$<\Delta_2(r_s)>\propto L^{-\alpha(r_s)}$, with $\alpha(r_s)$ 
going from $2$ to $1$ when $r_s$ goes from $0$ to $3$ approximately. 
The value $\alpha=2$ is expected without interaction 
($<\Delta_2>=<\Delta>\propto 1/L^2$). 
The exponent $\alpha=1$ can be understood as follows: for $r_s>1$, 
$\Delta_2$ is dominated by the charging energy and therefore 
$<\Delta_2>\propto 1/C\propto 1/L$.  
In the strongly localized regime ($W=15$) the $1/L$ law 
is justified due to the Coulomb gap in the single particle density 
of states \cite{ES}: According to Koopmans' theorem \cite{note}, 
assuming that the other charges are not reorganized by the addition 
of an extra charge, 
\begin{equation} 
\Delta_2\approx \epsilon_{N+1}-\epsilon_N\propto \frac{1}{L}  
\end{equation}
due to the Coulomb gap, where $\epsilon_k$ is the energy of the 
$k$-th HF orbital at a fixed number $N$ of particles.  

\begin{figure} 
\centerline{
\epsfxsize=8cm 
\epsfysize=8cm 
\epsffile{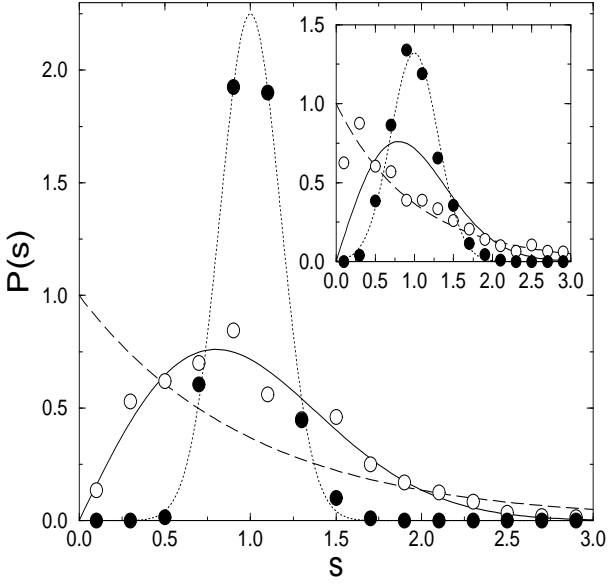}
}
\caption{ 
Distribution of the normalized inverse compressibilities 
for $N=16$, $L=12$, $W=5$, $r_s=0$ 
(empty circles) and $r_s=2.5$ (filled circles), fitted by a Gaussian 
of standard deviation $\sigma=0.18$ (dotted line). Solid and dashed 
lines give Wigner and Poisson distribution respectively.  
Inset: same at $W=15$; the Gaussian fit gives $\sigma=0.30$. 
}
\label{fig5} 
\end{figure}

 Fig. \ref{fig5} shows the distribution of inverse compressibilities, 
for $L=12$, $W=5$ and $W=15$ (inset). 
At $r_s=0$, $\Delta_2$ distributions are close to the Wigner surmise 
$P_W(s=\Delta_2/<\Delta_2>)=(\pi s/2)\exp(-\pi s^2/4)$ for 
$W=5$ and to the Poisson distribution $P_P(s)=\exp(-s)$ for $W=15$. 
In the latter case, deviations from the Poisson distribution at 
small $s$ values are due to the finite system size. 
In both cases at $r_s=2.5$ distributions show a Gaussian shape. 
This can be understood within the Koopmans' theorem, as the 
HF energies are given by 
\begin{equation}
\epsilon_k=\langle\psi_k| H_1 |\psi_k\rangle +\sum_{\alpha=1}^N 
\left(Q_{\alpha k}^{\alpha k}-Q_{\alpha k}^{k \alpha}\right),      
\end{equation} 
with $H_1$ one-body part of the Hamiltonian (\ref{hamiltonian}):  
Due to the small correlations of eigenfunctions in a random 
potential, one can reasonably invoke the central limit theorem
(see Ref. \cite{Berkovits} for a more detailed discussion).  

\begin{figure} 
\centerline{
\epsfxsize=8cm 
\epsfysize=8cm 
\epsffile{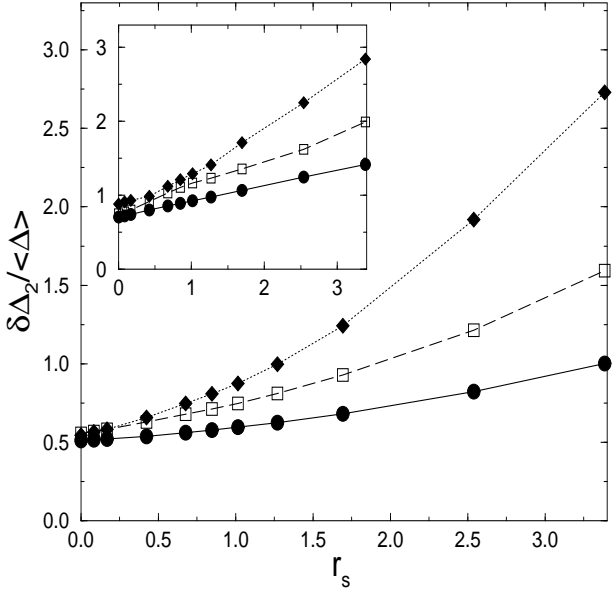}
}
\caption{ 
Inverse compressibility standard deviation $\delta\Delta_2$, normalized 
to the single particle mean level spacing $<\Delta>$ as a function of 
$r_s$, for $W=5$, $n_e=1/9$, $L=6$ (circles), $L=9$ (squares), and $L=12$ 
(diamonds). Inset: same at $W=15$.  
}
\label{fig6} 
\end{figure}

 Fig. \ref{fig6} shows the $r_s$-dependence of the inverse compressibility 
standard deviation $\delta\Delta_2$, normalized to the single particle 
mean level spacing $<\Delta>$.  
One can see that $\delta\Delta_2/<\Delta>$ increases slowly for $r_s<1$ and 
then more rapidly. This results are in contrast with random phase approximation
estimates \cite{BA,Blanter}, giving fluctuations of the order $\Delta$.  
On the contrary, we have checked that $\delta \Delta_2/<\Delta_2>$ has a 
weak $r_s$ dependence for $r_s>1$, in agreement with findings of 
Ref. \cite{Sivan}. This confirms that peak spacing fluctuations are 
dominated by capacitance fluctuations instead of single particle 
fluctuations.  

 We point out that our data are limited to spinless fermions. 
Interesting interaction-induced spin effects have been recently 
reported in experiments \cite{Ensslin} and theoretically 
investigated \cite{Brouwer,Baranger,Jacquod}.    

\section{First energy excitation} 
\label{gap} 

 We have calculated the first $N$-body energy levels $E_i$, $(i=0,1,2,...)$
for different sizes $L$, with a large disorder to hopping ratio 
$W/t=15$ imposed to have Anderson localization and Poissonian spectral 
statistics for the one particle levels at $r_s=0$ when $L\geq 8$. 
We studied $N=4,9$, and $16$ particles inside clusters of size $L=8,12$, 
and $16$ respectively. This corresponds to a constant low filling  
factor $n_e=1/16$. We studied an ensemble of $10^4$ disorder configurations.  

\begin{figure} 
\centerline{
\epsfxsize=8cm 
\epsfysize=8cm 
\epsffile{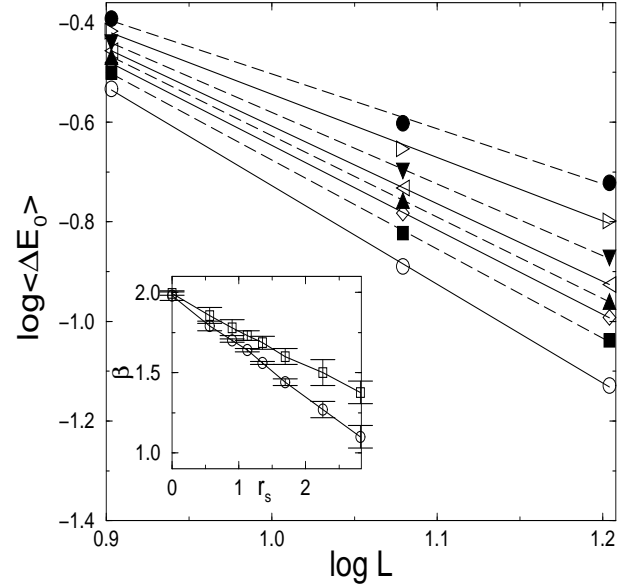}
}
\caption{ 
Size dependence of the average gap 
(first spacing $<\Delta E_0> \propto L^{-\beta (r_s)}$), 
for $W=15$, filling factor $n_e=1/16$. 
From bottom to top: $r_s=0,0.6,0.9,1.1,1.4,1.7,2.3,2.8$. 
Inset: $\beta (r_s)$ (circles, characterizing $<\Delta E_0>$ and 
squares, characterizing $<\Delta E_i>$, with an average over $i=1-3$).    
Disorder average is over $10^4$ configurations. 
}
\label{fig7} 
\end{figure}

 The first average spacing $<\Delta E_0>$ is given in Fig. \ref{fig7}. 
It exhibits a power law decay as $L$ increases, with an exponent $\beta$ 
given in the inset. 
One finds for the first spacing that $\beta$ linearly decreases from 
$d=2$ to $1$ when $r_s$ increases from $0$ to $3$. 
The next mean spacings $<\Delta E_i>=<E_{i+1}-E_i>$ depend more weakly 
on $r_s$, as shown in Fig. \ref{fig7}. 

\begin{figure} 
\centerline{
\epsfxsize=8cm 
\epsfysize=8cm 
\epsffile{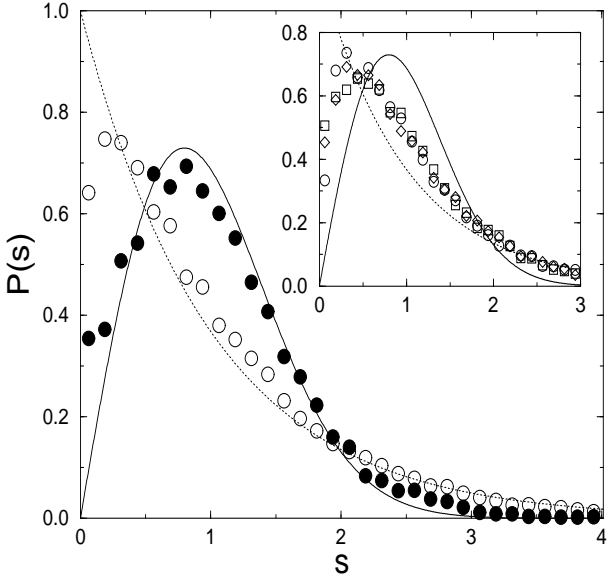}
}
\caption{Gap distribution $P(s)$ for $r_s=0$ (empty circles) and 
$r_s=2.8$ (filled circles) when $N=16$, $L=16$, $W=15$, compared to 
$P_P(s)$ and $P_W(s)$. Inset: size invariant $P(s)$ at 
$r_s^C\approx 1.2$ ; $L=8$ (circles), $12$ (squares), and $16$ 
(diamonds).
}
\label{fig8} 
\end{figure}

\begin{figure} 
\centerline{
\epsfxsize=8cm 
\epsfysize=8cm 
\epsffile{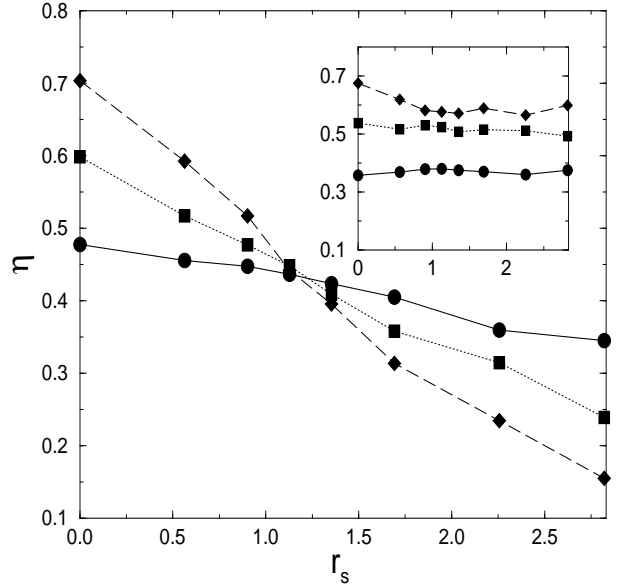}
}
\caption{ 
Parameter $\eta(r_s)$ corresponding to the first spacing $\Delta E_0$ 
at $n_e=1/16$, $W=15$, $L=8$ (circles), $12$ (squares), and $16$ 
(diamonds). Inset: $\eta(r_s)$ for the second spacing $\Delta E_1$. 
}
\label{fig9} 
\end{figure}

 For $r_s=0$, the distribution of the first spacing $s=\Delta E_0/<\Delta E_0>$
becomes closer and closer to the Poisson distribution $P_P(s)$ 
when $L$ increases, as it should be for an Anderson insulator. For a 
larger $r_s$, the distribution seems to become close to the Wigner surmise 
$P_W(s)$ characteristic of level repulsion 
in Random Matrix Theory, as shown in Fig. \ref{fig8} for $r_s = 2.8$ and 
$L=16$. 
To study how this $P(s)$ goes from Poisson to a Wigner-like distribution 
when $r_s$ increases, we have calculated a parameter $\eta$ which decreases 
from $1$ to $0$ when $P(s)$ goes from Poisson to Wigner:
\begin{equation} 
\eta=\frac{\hbox{var}(P(s))-\hbox{var}(P_W(s))} 
{\hbox{var}(P_P(s))-\hbox{var}(P_W(s))}, 
\end{equation}  
where $\hbox{var}(P(s))$ denotes the variance of $P(s)$, 
$\hbox{var}(P_P(s))=1$ and $\hbox{var}(P_W(s))=0.273$.
In Fig. \ref{fig9}, one can see that the three curves $\eta (r_s)$ 
characterizing the first spacing for $L=8, 12, 16$ intersect at 
a critical value $r_s^C \approx 1.2$. Our data suggest that 
for $r_s < r_s^C$ the distribution tends to Poisson in the thermodynamic 
limit, while for $r_s>r_s^C$ it tends to a Wigner-like behavior \cite{glass}. 
At the threshold $r_s^C$, there is a size-independent intermediate 
distribution shown in the inset of Fig. \ref{fig8}, exhibiting level 
repulsion at small $s$ followed by a $\exp( -a s)$ decay at large $s$ 
with $a \approx 1.52$. This Poisson-Wigner transition characterizes 
only the first spacing, the distributions of the next spacings being 
quite different. The inset of Fig. \ref{fig9} does not show an 
intersection for the parameter $\eta$ calculated with the second 
spacing. The second excitation is less localized than the first one 
when $r_s=0$, since the one particle localization length weakly increases 
with energy. It is only for $L=16$ that the distribution of the 
second spacing becomes close to Poisson without interaction, and a weak  
level repulsion occurs as $r_s$ increases. 

 The observed transition, and the difference between the first spacing and 
the following ones is mainly an effect of the HF mean field. For the first 
spacing, the curves $\eta$ calculated with the HF data are qualitatively the 
same. At the mean field level the low energy excitations are particle-hole 
excitations starting from the ground state. 
The energy spacing between the first and the second excited states is 
given by the difference of two particle-hole excitations and a 
Poisson distribution follows if the low energy particle-hole 
excitations are uncorrelated. 

 Notice that the energy of an electron-hole pair is given by 
$\epsilon_j-\epsilon_i-U/r_{ij}$ and the classical argument for the 
existence of a gap in the single particle density of states does not 
apply \cite{ES}. For example, in the constant interaction model one 
pays a gap $e^2/C$ in the single particle density of states but not 
in the many-body excitation spectrum ($\Delta E_0=\Delta E_0(U=0)
\propto 1/L^2$). Therefore the observed opening of a gap for the 
first energy excitation is a remarkable phenomenon beyond the 
predictions of the classical Coulomb gap model.  

\section{Hopping conductivity}
\label{hop} 

 In disordered insulators, a crossover \cite{ES} in the temperature 
dependence of the resistivity $\rho(T)$ is induced by Coulomb repulsion  
from the Mott variable range hopping law 
($\rho(T)=\rho_M \exp (T_0/T)^{1/3}$ in $2d$) 
to the Efros-Shklovskii behavior 
($\rho(T) = \rho_{\rm{ES}} \exp (T_{\rm{ES}}/T)^{1/2}$). 

 The usual argument is to consider the length $L(T)$ which maximizes   
the probability $P_{ij}$ of a single particle hop from a state with energy 
$\epsilon_i$ to a state with energy $\epsilon_j$:   
\begin{equation} 
P_{ij}\propto\exp\left(-\frac{2L}{\xi}-\frac{\epsilon_j-\epsilon_i}{kT}\right), 
\end{equation} 
with $\xi$ localization length.

 To measure possible delocalization effects, we have calculated for 
$W=15$ the number of occupied sites per particle 
$\xi_s=N/\sum_{i} \rho^{2}_i$, where $\rho_i=\langle\Psi_0|n_i|\Psi_0\rangle$ 
is the charge density of the ground state at the site $i$.  
Around $r_s \approx 1.2$ and after ensemble average, the maximum increase 
of $\xi_s$  compared to $r_s=0$ is negligibly small ($2 \%$). 
Therefore noticeable effects come mainly from the mean and the distribution 
of the single particle density of states around the Fermi level. 

 At low temperatures, hopping conductivity involves states near 
the Fermi level. If one takes for the Coulomb gap its average 
value $\epsilon_{N+1}-\epsilon_N \approx \Delta_2 \approx (A+B r_s)/
L^{\alpha(r_s)}$ (see Fig. \ref{fig4}), one obtains for the hopping 
resistivity a smooth and continuous crossover from Mott to Efros-Shklovskii 
hopping, given by: 
\begin{equation}
\rho(T)\propto \exp \left(\frac{T(r_s)}{T}\right)^{1/(\alpha+1)}, 
\end{equation} 
where 
\begin{equation}
T(r_s) \approx \frac{A+B r_s}{k \xi^{\alpha}}.
\end{equation}
This prediction neglects the crossover from the Poisson distribution to 
a Gaussian distribution in the inverse compressibility (see Fig. \ref{fig5}), 
which could be included by considering for $\Delta_2$ a more typical value 
than its average, for instance the value $\Delta_2(b)$ for 
which $\int_0^{\Delta_2(b)} P(\Delta_2^{'}) d\Delta_2^{'}=b$, with $b=1/2$ 
for example. This would introduce a crossover in $T(r_s)$ around $r_s\approx 1$. 

 The existence of a critical $r_s$ value for the opening of the Coulomb
gap can be understood similarly to \cite{PD99}. The single particle 
density of states around the Fermi energy $E_F$ is given by 
\cite{ES} $\rho(E)\approx |E-E_F|/U^2$ and the gap size 
$\Delta_g=|E_g-E_F|$ can be estimated from the condition 
$\rho(E_g)\approx\overline{\rho}$, 
with $\overline{\rho}\approx 1/W$ mean density of states for $W\gg t$, 
obtaining $\Delta_g\approx U^2/W$.  
According to Fermi golden rule, the inverse lifetime of a Slater 
determinant built from electrons localized at given sites is   
$\Gamma_t\approx t^2 (1/W) (N/L^2)$, with $N/(W L^2)$ density 
of states directly coupled by the hopping term of the Hamiltonian 
(\ref{hamiltonian}). 
This argument suggests that quantum fluctuations can melt the Coulomb 
gap for $\Gamma_t\approx\Delta_g$, giving $r_s\approx 1$. 
Therefore a crossover from Efros-Shklovskii to Mott hopping conductivity 
is expected not only increasing temperature but also increasing carrier 
density, as observed in \cite{khondaker} at $r_s\approx 1.7$.   

 Notice that in the above arguments we have considered only the excitations 
leading to the Efros-Shklovskii behavior for the hopping conductivity, that 
correspond to the withdrawal of one electron or its addition to the system. 
However, a single electron hop may reorganize the location of the other 
particles, inducing complex many particle excitations (see Fig. \ref{fig2}, 
showing that simple HF is unable to reproduce the mean first many-body 
energy spacing when $r_s>1$). Therefore one cannot exclude that correlated 
hops involving more than one particle affect the hopping conductivity 
\cite{ES}. 

\section{Conclusions} 
\label{conc} 
 
 In summary, we have analyzed the inverse compressibility and the first energy 
excitation statistics for spinless fermions in weakly and strongly disordered 
squared lattices when $r_s$ increases from $0$ to $3$. On one hand, we have 
found a crossover in the electronic inverse compressibility from a $1/L^2$ 
towards a $1/L$ decay; at the same time its distribution evolves towards a 
Gaussian shape. On the contrary, our data for the first many-body energy 
spacing are consistent with a smooth opening of a gap and a sharp 
interaction-induced Poisson-Wigner-like transition in its distribution 
at $r_s^C\approx 1.2$, with a scale invariant distribution at the critical 
point showing intermediate statistics.  For those small values of $r_s$, 
the ground state energy is essentially given by an effective 
Hartree-Fock mean field. We underline that the intermediate quantum regime 
discussed in Ref. \cite{paper1} and charge crystallization occur 
for larger $r_s$ factors, where the mean-field approximation 
breaks down and the configuration interaction method ceases to 
efficiently converge. This does not allow us to study the regime 
characterizing the experiments reported in Refs. \cite{comp1,comp2}.

\begin{acknowledgement} 
Partial support from the TMR network ``Phase Coherent Dynamics of 
Hybrid Nanostructures'' of the European Union is gratefully acknowledged.  
\end{acknowledgement}

\end{document}